\documentclass[iop]{emulateapj}
\usepackage{color}
\usepackage{natbib}
\slugcomment{}

\shorttitle{M54+Sgr=$\omega$ Cen}
\shortauthors{Carretta et al.}

\begin{document}
\title{M~54 $+$ Sagittarius = $\omega$ Centauri \altaffilmark{1}}

\author{E. Carretta\altaffilmark{2}, A. Bragaglia\altaffilmark{2},
R.G. Gratton\altaffilmark{3},
S. Lucatello\altaffilmark{3,4},
M. Bellazzini\altaffilmark{2},
G. Catanzaro\altaffilmark{5},
F. Leone\altaffilmark{6},
Y. Momany\altaffilmark{3,7},
G. Piotto\altaffilmark{8},
V. D'Orazi\altaffilmark{3}}

\altaffiltext{1}{Based on data collected at the ESO telescopes under 
programme 081.D-286}
\altaffiltext{2}{INAF, Osservatorio Astronomico di Bologna, via Ranzani 1,
       40127,  Bologna,  Italy. eugenio.carretta@oabo.inaf.it,
       angela.bragaglia@oabo.inaf.it, michele.bellazzini@oabo.inaf.it}
\altaffiltext{3}{INAF, Osservatorio Astronomico di Padova, vicolo
       dell'Osservatorio 5, 35122 Padova,  Italy. raffaele.gratton@oapd.inaf.it
       sara.lucatello@oapd.inaf.it, valentina.dorazi@oapd.inaf.it}
\altaffiltext{4}{Excellence Cluster Universe, Technische Universit\"at M\"unchen, 
Boltzmannstr. 2, D-85748, Garching, Germany}
\altaffiltext{5}{INAF-Osservatorio Astrofisico di Catania, Via S.Sofia 78, I-95123 
Catania, Italy}
\altaffiltext{6}{Dipartimento di Fisica e Astronomia, Universit\`a di Catania, Via S.Sofia 78, I-95123 
Catania, Italy}
\altaffiltext{7}{European Southern Observatory, Alonso de Cordova 3107, Vitacura, 
Santiago, Chile}
\altaffiltext{8}{Dipartimento di Astronomia, Universit\`a di Padova, Vicolo dell'Osservatorio 2,
I-35122 Padova, Italy}

\begin{abstract}

We derive homogeneous abundances of Fe, O, Na and $\alpha-$elements from high
resolution FLAMES spectra for 76 red giant stars in NGC~6715 (M~54) and for 25
red giants in the surrounding nucleus of the Sagittarius (Sgr) dwarf galaxy. Our
main findings are that: (i) we confirm that M~54 shows intrinsic metallicity
dispersion, $\sim 0.19$ dex r.m.s.;
(ii) when
the stars of the Sgr nucleus are included, the metallicity distribution strongly
resembles that in $\omega$~Cen; the relative contribution of the most metal-rich
stars is however different in these two objects; (iii) in both GCs there is a
very extended Na-O anticorrelation, signature of different stellar generations
born within the cluster, and (iv)  the metal-poor and metal-rich components in
M~54 (and $\omega$ Cen) show clearly distinct extension of the Na-O
anticorrelation, the most heavily polluted stars being those of the metal-rich
component. 
We propose a tentative scenario for cluster formation that could explain these features.
Finally, similarities and differences found
in the two most massive GCs in our Galaxy can be easily explained if they are
similar objects (nuclear clusters in dwarf galaxies) observed at different
stages of their dynamical evolution. 

\end{abstract}


\section{Introduction}

The complex, multi-population nature of globular clusters (GCs) is presently
well assessed (see \citealt{gra04} for an extensive  review): there is evidence
of multiple sequences from photometry in some GC  and more universally from
spectroscopy (see e.g. \citealt{pio09}). Our survey of 19 GCs
(\citealt{car09a,car09b})
aconfirmed that star-to-star variations in light elements
(e.g., O, Na, Mg, Al, Si) are present in all GCs studied; this pattern was
produced by the ejecta of now extinct more massive stars (\citealt{gra01})
through H-burning at high temperature (\citealt{den89,lan93}).
However, these  variations in light elements do not generally extend
to heavier elements: the upper limit we found for the scatter in [Fe/H] is less
than 0.05 dex (i.e., the homogeneity is better than 12\%, \citealt{car09c})
On the other hand, {\em intrinsic} spread in Fe abundances has been
found in a few GCs, including $\omega$ Cen 
(\citealt{fre75,but78,nor95} - NDC95 - and several other studies; see
\citealt{gra04} for extensive references), and the recently scrutinized
NGC~6656 (M~22, \citealt{mar09}), Terzan~5 (\citealt{fer09}), and
possibly M~54 (\citealt{sar95,bel08}: B08, and
references  therein).

The most striking signal of the early pollution is the Na-O anticorrelation.
While its shape may differ from cluster to cluster, its widespread existence led
us to associate it to the same definition of GC (\citealt{car09d}). We need
an adequate sampling of the Na-O anticorrelation to shed light on the complex
scene of the initial cluster evolution, likely occurring in the core of giants
clouds/associations or even in the core of dwarf 
galaxies (\citealt{bek07,bok09}).

Combining information coming from the chemistry and the color-magnidude diagrams
(anticorrelations between elements, inferred He enhancement, multiple/complex
main sequences and subgiant branches, horizontal branches, see \citealt{bra10}
for a recent review), we may be able to put together several pieces of the
puzzle and reach a more in-depth understanding of star formation in dense
environments. This will offer a circumstantiated answer to fundamental questions
such as how the GCs formed and whether they were able to build at least part
of their metals.

In this context, M~54 (NGC~6715) appears as a key object. It is an  old,
metal-poor (e.g., \citealt{lay98}), massive GC immersed in the
nucleus of the Sagittarius dwarf spheroidal (Sgr dSph) galaxy, presently disrupting
within our Galaxy (\citealt{iba94}, B08, and
references therein). M54 is the most massive of the four GCs
associated to the Sgr dSph, and it has a very extended horizontal branch (HB)
with a population of ``blue hook" stars, found only in a few of the most massive
GCs (\citealt{ros04}). Therefore,  it is at the same time the nearest {\it bona
fide} extragalactic cluster and the second most massive GC in the Milky Way,
after $\omega$ Cen (\citealt{har96}). 

M~54 and $\omega$~Cen represent the high mass tail of the GC mass distribution,
and show many similarities, worth of deeper insight: i) both have intrinsic
dispersion in metallicity [Fe/H], even if of different amplitude; ii) they are
either associated to (M~54) or suspected to be born in ($\omega$ Cen) a dwarf
galaxy; iii) both lie in the intermediate region (in the $M_V$ vs half mass
radius) between Ultra Compact Dwarfs (UCDs) and GCs, very close to the low-mass
limit of the UCDs (see Fig. 1 in \citealt{tol09,mac05,fed07}).
All these similarities may led to the legitimate suspicion that M~54 and
$\omega$ Cen could be siblings, or at least next of kin. In particular B08
speculated that the residual of the (future) complete dissolution of the Sgr
galaxy will leave a long-living compact remnant composed by a bulk of metal-poor
stars (the original M54 cluster) plus a lesser population of metal-rich stars
from the original nucleus of Sgr (Sgr,N). This would be very similar to the
current status of $\omega$ Cen (see \citealt{pan00}, and references
therein), suggesting that this puzzling system may have formed through an analogous
process.

In this paper we investigate this scenario in further detail, comparing newly
obtained abundance analysis from high dispersion spectra of M54 (and Sgr,N) 
stars with similar data for $\omega$~Cen taken from the literature. 

\section{Spectroscopic data}

\subsection{M54}

The only previous study of M~54 based on high resolution spectroscopy was that
of \citet[hereinafter BWG99]{bro99} who analyzed five giant stars. They found an
average Fe/H]$=-1.55$ dex and evidence of proton-capture reactions (low O,
enhanced Na and Al) in the abundance ratios of one, perhaps two stars. 

Our study is based on the FLAMES (GIRAFFE and UVES) spectra of 76 stars on the
red giant branch (RGB) of M~54, and of 25 stars belonging to the Sgr dSph
nucleus; the two samples are selected from the RGBs of the two populations that
are well separated in the CMD (B08). 
A full description of the analysis and results will be presented elsewhere
(Carretta et al. 2010, in preparation). In this Letter we only show results
concerning Fe, Na, O and the two $\alpha-$elements Mg, Si. However, our abundance
analysis traces as closely as possible the homogeneous procedures adopted for
other GCs (see \citealt{car09a,car09b}  and references therein)\footnote{For 
instance, the temperature is derived from a relation between $T_{\rm eff}$ (from
$V-K$ and the \citealt{alo99} calibration) and $K$ magnitudes, much more
reliably measured than colours. This  leads to very small internal errors in the
atmospheric parameters, and hence in the derived abundances. This approach is
not valid for the Sgr dSph stars: in this case the adopted temperatures are simply
those from $V-K$ colours.}.

\subsection{$\omega$~Cen}

$\omega$~Cen has been extensively studied at high spectral resolution. We
consider here the following data sets:
\begin{itemize}
\item Metallicity distribution: we used data from \citet{sta07}, which
is the most recent large sample (about 380 stars) with metallicity publicly
available in literature, plus  the about 20 stars in \citet[O03]{ori03},
the only spectroscopic  study including the metal-rich branch RGB-a;
\item Na-O anticorrelation: to date, the only analysis of $\omega$~Cen
comparable to the present one for M~54 is the extensive data set from
NDC95, who obtained homogeneous abundances from high resolution spectra for 40
RGB stars\footnote{A similar, but more extensive analysis is corrently under
way, and preliminary results seem to support the findings of the present
Letter, A.F. Marino (2009, private communication).}.
\end{itemize}

\section{The comparison between M54 and $\omega$~Cen}

\subsection{Metallicity dispersion}

The average metallicity for M~54 derived from neutral Fe lines is
[Fe/H]$=-1.559\pm 0.021$~dex ($\sigma=0.189$~dex, 76 stars), which is almost
coincident with the value obtained by BWG99. We obtained an average [Fe/H] value
of $-0.622\pm 0.068$ dex, $\sigma=0.353$~dex for the 25 stars of the Sgr dSph
nucleus. Since internal errors in [Fe/H]{\sc i} in our analysis are $\sim
0.02$~dex, {\em we confirm (at more than 8$\sigma$) the existence of a
metallicity dispersion in M~54}, as proposed by \citet{sar95} and
B08.

\begin{figure} 
\plotone{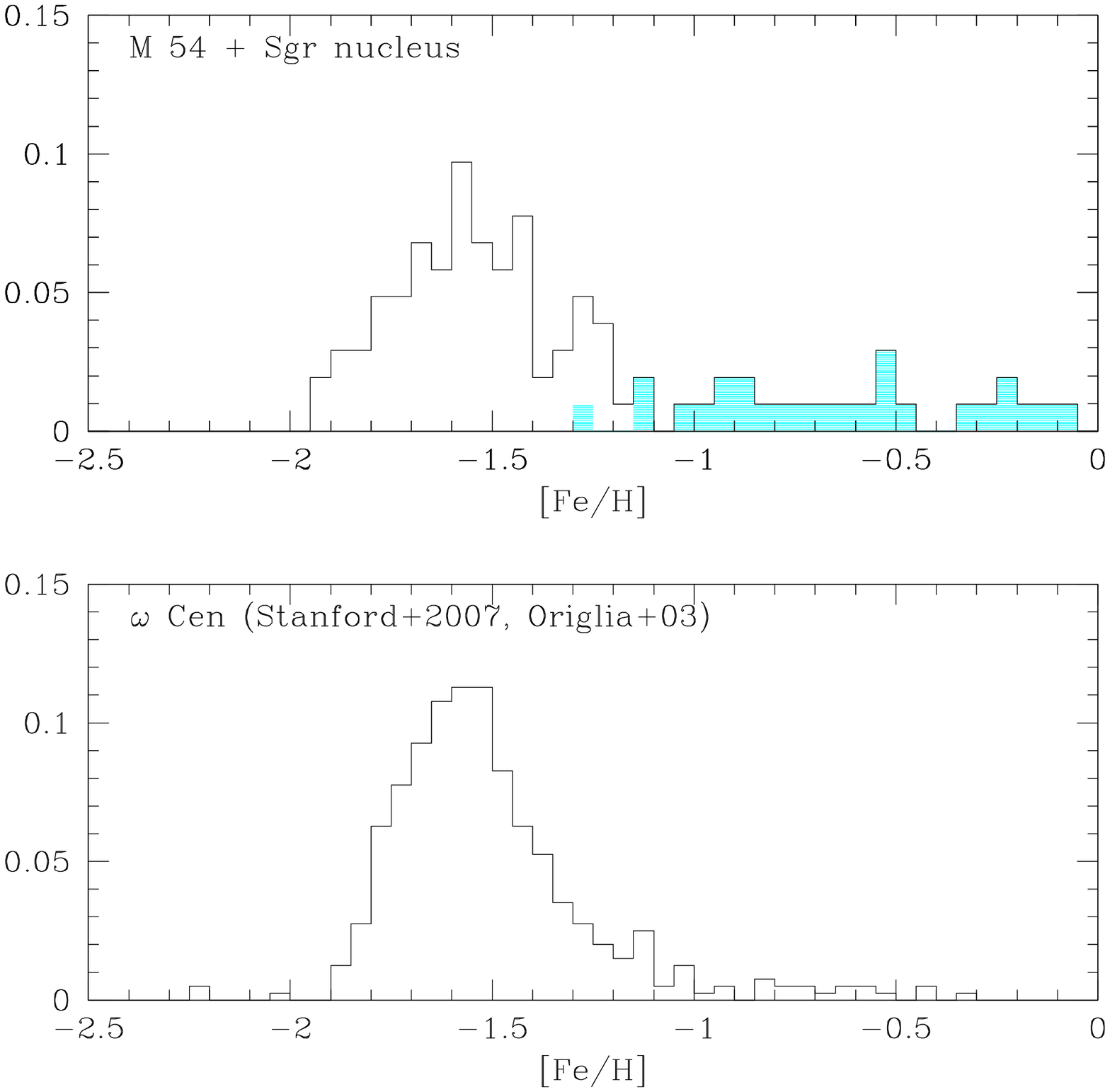}
\caption{Metallicity distribution function (MDF) of our total sample 
(M~54$+$Sgr nucleus -the latter also evidenced in color) as compared to the 
same for $\omega$ Cen from a sample of about 380
unevolved stars by \citet{sta07} plus the about 20 giants by
\citet{ori03}.  Both distributions are normalized to the total
number of objects in the sample.}
\label{f:metM54SGRN51}
\end{figure}

In Fig.~\ref{f:metM54SGRN51} we compare metallicity distribution functions (MDF)
in M~54$+$Sgr dSph nucleus (our data) and $\omega$ Cen (\citealt{sta07}, 
O03).
MDFs are formally different, however, similarities may be traced in the global
appearence. In both cases:
\begin{itemize}
\item the bulk of stars is metal-poor, with a major peak at 
[Fe/H]$\sim -1.6 \div -1.5$ dex; 
\item this peak is followed by a gradual decrease toward increasing metallicity,
up to [Fe/H]$=-1.0$ with more or less evident secondary peaks; 
\item a long tail up to solar metallicities is observed.
\end{itemize} 

The same conclusions are also evident for $\omega$ Cen in Fig. 9 
of \citet{nor96}.
While the exact relative fraction of the various component is obviously affected
by different selection effects acting in the different samples, the similarity
of the overall shape is well established and intriguing.

\subsection{The Na-O anticorrelation}

The Na-O anticorrelations obtained for M~54 (our data) and for $\omega$~Cen
(NDC95) are shown in Fig.2. They are the two most pronounced
known examples of O-depletions anticorrelated with Na-enhancements among RGB
stars in GCs. The interquartile range of the distribution of [O/Na] ratios,
assumed as a quantitative measure of the  extension of the anticorrelation (see
\citealt{car06}) are IQR[O/Na]=1.169 and 1.310 for M54 and $\omega$ Cen,
respectively. While the Primordial component fractions ($26\pm 6\%$\ 
and $31\pm 9\%$, respectively: see definition in \citealt{car09a}) are 
very similar to the average fraction of first generation stars in other GCs
(about 33\%), M~54 has the highest fractions of second generation stars with
extreme composition found up to date: E$_{\rm M~54}$ = $28\% \pm 6\%$.  The
values of IQR easily exceeding the previous record detained by NGC~2808
(\citealt{car09a}), M~54 and $\omega$~Cen nicely extend to the most massive
clusters in the Galaxy the tight correlation between the extension of the Na-O
anticorrelation and the total cluster mass (\citealt{car09d}). This
provides an indirect support to the very same definition of GC as given in
\citet{car09d}: in a {\it bona fide} GC the Na-O anticorrelation is
$always$ observed, whether its luminosity is faint like that of NGC~6838 (M~71,
$M_V=-5.60$, \citealt{har96}) or comparable to that of the faintest dwarf galaxies,
like the cases of M~54 ($M_V=-10.01$, $ibidem$) and $\omega$~Cen ($M_V=-10.29$,
$ibidem$).

However, our most striking finding is that {\it in both M~54 and $\omega$~Cen
this pattern has a clearly different extent if we regard separately the
metal-poor and metal-rich components} (separated at [Fe/H]$=-1.56$ dex, as shown
in the middle and right panels of Fig.~2 - this is of course a simplification): the metal-rich component reaches
higher degrees of processing by proton-capture reaction in H-burning at high
temperature\footnote{Already NDC95 (p. 695) noticed an {\it
apparent dependence of the operation of the ON cycle on [Fe/H]}, confirming
early findings by \citet{coh86} and \citet{pal89}.}. 
This difference is exactly mirrored also in the outcome of the Mg-Al cycle
(Carretta et al. 2010, in preparation) using our data for M~54 and using the
data from NDC95 and from intermediate resolution spectroscopy of a quite large
sample of giants with Na and Al abundances recently derived by \citet{joh09}
in $\omega$ Cen. A metallicity dependence of the distributions of Na and
Al was already noted by the latter authors.

In summary, we found that the two most massive GCs in the Galaxy have a very
similar pattern of anticorrelations and correlations for proton-capture
elements. We suggest that this is not due to a mere chance, but it is rather a
consequence of the way these large GCs formed.

\subsection{The metal-rich nuclear component(s)}

What about the stars of the Sgr dSph nucleus? The large spread in O and Na
abundances (anticorrelated with each other) is confined only to stars in M~54.
Those in Sgr dSph nucleus present a run of these elements as a function of the
metallicity typical of Galactic field stars, apart from the well known offsets
already observed for stars in Sgr dSph (e.g. \citealt{sme02,sme09,sbo07}).
This suggests that these stars did not participate to the episode of formation
of M~54, as also proposed on different grounds by B08.

There is not (any more?) a dwarf galaxy around $\omega$ Cen. However, in
Sect.3.1, by comparing the MDFs, we anticipated the idea
that the two most massive globular clusters observed in our Galaxy might have
followed a very similar evolutionary path, with M~54 being seen still
``frozen" in a earlier phase with respect to $\omega$ Cen.  In the case of M54, the
population belonging to the Sgr dSph is prominent at high metallicity. For
$\omega$~Cen only few studies insofar were focused on the most metal-rich
population, apart from   O03 that analyzed\footnote{Or
re-analyzed stars from \citet{pan03}.} a few stars of the so called
anomalous RGB (RGB-a) using infrared low resolution spectroscopy.

In Fig.3 we compare our results for M~54 plus Sgr dSph (details will be presented
in Carretta et al. 2010, in preparation) with those for $\omega$~Cen (NDC95) including stars
on the RGB-a (O03). While offsets between infrared and optical spectroscopy might well be
present, it is quite evident that stars on the most metal-rich RGB in
$\omega$~Cen show the same lack of anticorrelation between p-capture elements
observed for stars in the Sgr dSph nucleus. Again, the similarity of the
trends is striking and suggests that in the case of $\omega$ Cen {\it the most
visible residual of the ancestral galaxy once surrounding this cluster is 
probably represented by the so-called RGB-a.}\footnote{There are also 
indications, though debated, of
different proper motions and ages, for stars on the different branches of the
CMD of $\omega$ Cen, see \citet{gra04} for references.} 

Searches for likely debris of $\omega$ Cen proto-galaxy are actively ongoing
(e.g. \citealt{wyl09,mez05,miz03}, D.
Romano, private communication), but a residual is almost certainly under our
eyes since many years, still linked to the cluster itself, as a witness of the
same process of nucleation still at work in the case of M~54 (B08).

\section{Discussion and conclusions}

We found intriguing analogies between the chemical composition of stars in M~54
and in $\omega$~Cen. This supports the idea that these two objects (the most
massive GCs of the Milky Way) may have formed in a very similar way, and
represents just two subsequent snapshots of the same basic evolution of 
dwarf (nucleated?) galaxies,
taken at different times. The more advanced
stage reached by $\omega$~Cen (completely dissolved parent galaxy) is expected,
as the peri-Galactic of its orbit is much closer to the Galactic center than
that of Sgr+M54, hence it should have suffered much stronger tidal stresses.

Coming back to the formation phase of these massive GCs, is it possible to
include them in the scenario for GC formation proposed by \citet{car09d}?
This scenario proposes that formation of GCs started from strong
interactions of cosmological large fragments either among them or with the main
Galaxy. This strong interaction started the transformation of some gas in stars, 
forming a $precursor$ population whose core-collapse SNe homogeneously
enriched in metals the system and further triggered a large burst of star
formation. This massive episode (corresponding to the formation of the $primordial$ population)
lasted until the primordial gas was swept away by winds of SN II and
massive stars. The sudden mass loss caused expansion of the
primordial stars as an unbound association. However, just after this phase of
energetic winds, low velocity winds from evolving fast rotating massive stars or
intermediate-mass AGB stars replenished (or re-collected via a cooling flow) a
new kinematically cold gas reservoir in the central regions of the association,
where second generation stars may form within a very compact central cluster
(the one we are currently observing as a GC and which contains a dominant fraction
of second generation stars).

The results for M~54 and $\omega$ Cen might be explained by a tentative
extension of this scenario. First, \citet{geo09}
noted that M~54 and $\omega$ Cen are located in the same region of the mass-half
mass radius plane populated by nuclear star clusters in dwarf galaxies. However,
both clusters are shifted to larger sizes. They explain this observation as a
consequence of an early expansion of the GCs due to an abrupt change of the 
potential following a strong interaction of their parent system with the main
Galaxy. Second, we attribute the $very$ metal-rich tail, which 
does not show a Na-O anticorrelation, to accretion of stars, either younger or older,
from the parent galaxy;  they did not
participate to the formation of these massive GCs, and we will then neglect
them. 

For what concern the dominant cluster populations, the truly intriguing fact is
that the anticorrelation is more extended at higher metallicities, where we
observe a large incidence of extremely O-poor, and likely very He-rich stars.
In fact, we expect the extension of the Na-O anticorrelation to be
determined by the range of masses of the polluters. Taking the
AGB case, larger mass polluters (6-8
$M_\odot$, lifetimes $\sim 40-70$ Myr, using the isochrones by \citealt{mar08})
produce an extended anticorrelation, with very low O abundances, and large
production of He, while smaller mass polluters (5-6 $M_\odot$, lifetime $\sim 70-100$
Myr) produce a much less extended anticorrelation, with  minimal effects on He.

The explanation we propose for our observations requires appropriate
geometry and timing: the metal-poor second generation needs to be formed
by the ejecta of $\sim 5-6~M_\odot$\ AGB stars, and the metal-rich one
{\it only} by the ejecta of $\sim 6-8~M_\odot$\ AGB's. If we want to avoid
the contribution by the most massive polluters to the metal-poor second
generation, we need a delay of the cooling flow (or gas replenishing) from
this population by $\sim 10-30$~Myr. This can be obtained if we assume
that the metal-rich component formed $\sim 10-30$~Myr {\it later} than the
metal-poor one from material further enriched in metals (a reasonable but
not demonstrated assumption). The easiest way to get this is to have two
(or more) close but distinct regions (this may be expected if the region
of star formation is very large). Let us call A and B these regions. With
this small time offset, while the $\sim 6-8~M_\odot$\ stars of the
metal-poor component (A region) are in their AGB phase, massive stars of
the metal-rich component (B region) are still exploding as core-collapse
SNe. The large kinetic energy injected into both A and B regions prevents
the gathering of gas from both metal-poor and metal-rich populations. In
this phase there is not any cooling flow yet. When the rate of SN
explosions in region B becomes low enough, a quiet phase follows, lasting
some tens Myr. In this phase, cooling flow formation is possible for both
the metal-poor component (region A, with AGB polluters of $\sim
5-6~M_\odot$) and for the metal-rich one (region B, AGB polluters of $\sim
6-8~M_\odot$). In this scenario the second generation stars form nearly
simultaneously in the metal-poor (region A) and metal-rich (region B)
cooling flows. Finally, the SN explosions of stars formed in these cooling
flows (or the onset of type Ia SNe) stop this later phase of star
formation. The initially binary (or multiple) protocluster has then ample
time to merge for dynamical friction, and to appear as a single system,
but with complex chemical composition. Of course, in the case of
$\omega$~Cen, we may have more than two regions.

In this tentative hypothesis, the main difference between typical and
massive GCs is that in the latter the star formation continues at a high rate
for a more prolonged period (although shifting toward other regions of the global
star forming area) than in the case of the small mass clusters. 
Of course, this scenario needs verifications, and it is for the moment quite
speculative. However, we think it represents a reasonable extension to larger
masses of what we have considered insofar for more typical GCs.
In any case, the case presented here offers further support to the connection between 
GCs and dwarf galaxies (see also \citealt{fre02} on a model for $\omega$ Cen 
similar to the one proposed here).

\acknowledgements
We warmly thank Andy McWilliam for sending us his manuscript and data on Sagittarius in
advance of publication. EC wishes to thank Lucia Ballo for useful comments and
suggestions. Partial funding come from the PRIN MIUR 2007 CRA 1.06.07.05, PRIN INAF 2007 CRA 1.06.10.04,  the DFG
cluster of excellence ''Origin and Structure of the Universe''.

\clearpage

\clearpage

\clearpage

\begin{figure} 
\plotone{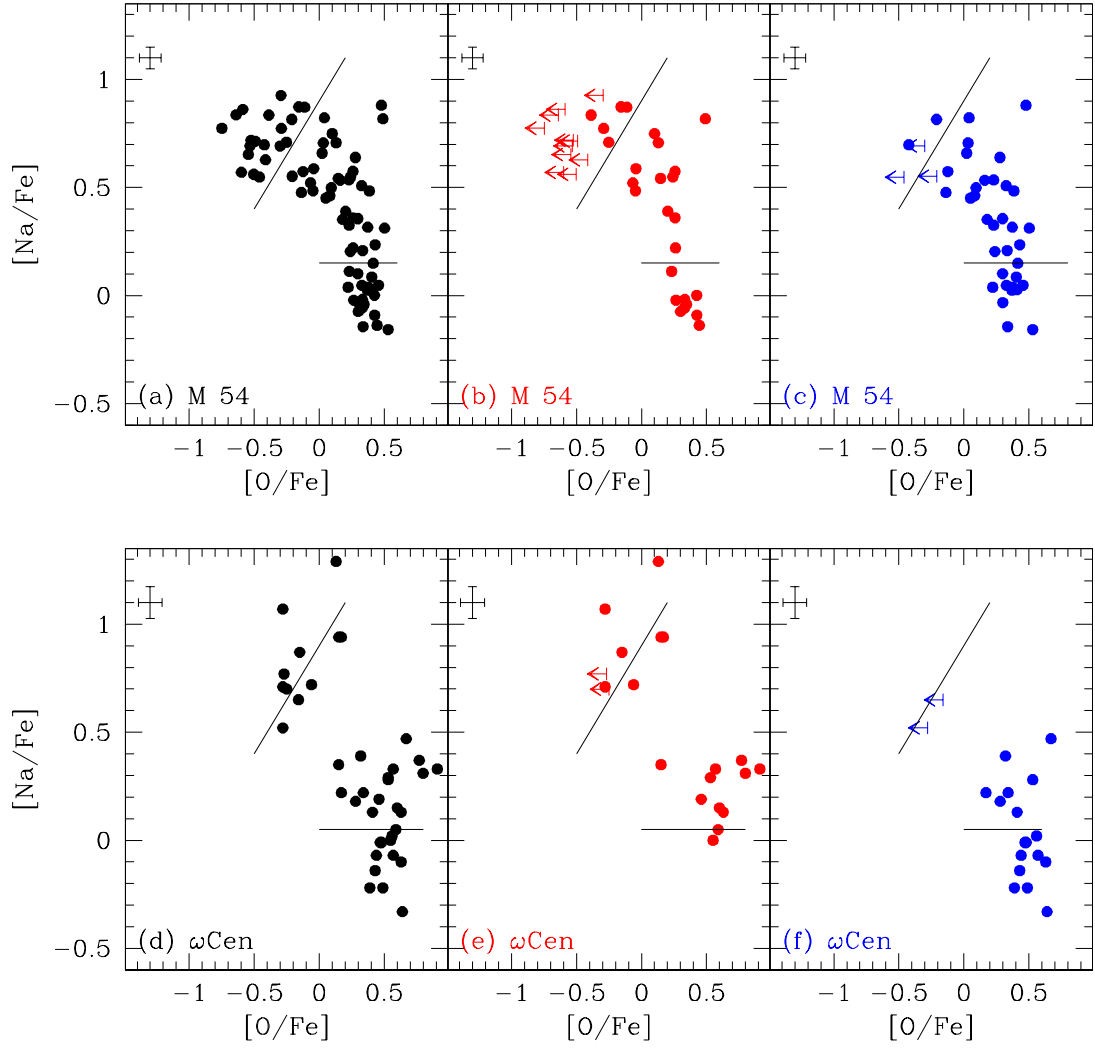}
\caption{(a) Na-O anticorrelation in M~54 from both UVES and GIRAFFE
spectra.(b), (c) The same, separating stars that are more metal-rich and
more  metal-poor than the cluster average [Fe/H]$=-1.56$, respectively. Upper
limits in [O/Fe] are indicated by arrows and the typical star-to-star error bars
are also shown. (d), (e), (f) The same, for $\omega$ Cen, from NDC95.} 
\label{f:figanti}
\end{figure}

\clearpage

\begin{figure} 
\plotone{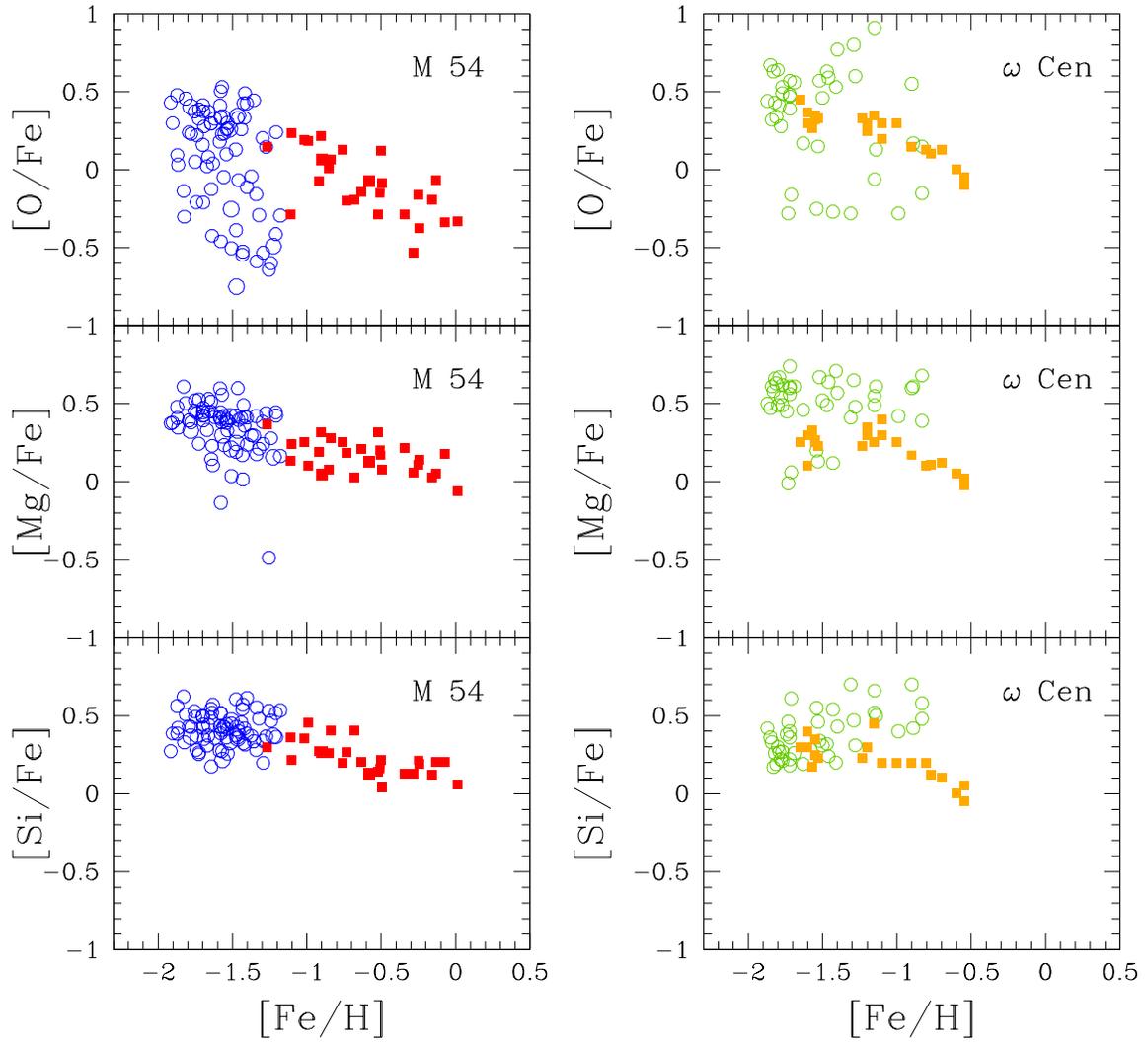}
\caption{Left panels, from top to bottom: [O/Fe], [Mg/Fe] and [Si/Fe] ratios  as
a function of [Fe/H] from our analysis of M~54 (circles) and Sgr nucleus
(squares). Right panels: the same for $\omega$ Cen using data by NDC95 (circles)
and from O03 (squares). The last sample includes stars on the RGB-a.}
\label{f:rgba8}
\end{figure}

\end{document}